\documentclass[final,5p,times,twocolumn]{elsarticle}

\usepackage{amssymb}
\usepackage{lipsum}
\usepackage{amsmath}
\usepackage{adjustbox}
\usepackage{comment}
\usepackage{subcaption} 
\usepackage{float}      
\usepackage{graphicx}
\usepackage{soul}
\sethlcolor{yellow} 
\usepackage{color}

\begin{document}

\begin{frontmatter}



\title{Epidemic Extinction in a Continuous SIRS Model with Vaccination}


\author[first]{Germano Hartmann Brill, Pablo Enrique Jurado Silvestrin, Sebastian Gonçalves}
\affiliation[first]{organization={Instituto de Física, Universidade Federal do Rio Grande do Sul},
            addressline={}, 
            city={Porto Alegre},
            postcode={Caixa Postal 15051, 91501-970}, 
            state={RS},
            country={Brazil}}

\begin{abstract}
Epidemics have shaped human history, often with devastating consequences, motivating the development of mathematical models to understand and control their dynamics. Among the many aspects of epidemic behavior, the conditions that lead to epidemic extinction stand out as a central---if not the fundamental---question in epidemic modeling. In this work, we study epidemic extinction in a continuous SIRS (Susceptible–Infected–Recovered–Susceptible) model governed by a system of ordinary differential equations (ODEs). The model includes vaccination as a time-dependent process and considers the reinfection of recovered individuals through waning immunity. We analyze how different parameter regimes---particularly infection, recovery, and immunity loss rates---affect the persistence or extinction of the epidemic. Special attention is given to the limitations of continuous population models, in which the infected fraction can fall below the equivalent of a single individual, leading to nonphysical outcomes such as unrealistically long persistence or artificial secondary peaks. By comparing the continuous SIRS dynamics with expected real-world thresholds for extinction, we highlight the importance of incorporating stochasticity or discrete effects to accurately describe epidemic fade-out.

\end{abstract}




\end{frontmatter}




\section{Introduction}
\label{introduction}

Epidemics of infectious diseases have occurred throughout human history, causing an enormous number of deaths. Smallpox alone was responsible for between 300–500 million deaths during the twentieth century~\cite{holmes2017major}, while the COVID-19 pandemic killed about 7 million people in the first three years~\cite{who2026coviddeaths, worldometer2025covid}. These impressive figures highlight how essential epidemic modeling has become to help guide public health strategies.

The primary focus of this article is the SIRS (Susceptible–Infected–Recovered–Susceptible) model with vaccination. This article examines the extinction in the SIRS model via Ordinary Differential Equations (ODEs). This model considers a continuous population partitioned into 3 groups: the susceptible $(S)$, the infected $(I)$ and the removed $(R)$. The population is considered constant, not accounting for births and deaths \cite{Hethcote2000, KeelingRohani2008, murray_mathematical_1993}, and is also normalized in order to represent the fractions of $S$, $I$ and $R$, as shown in Eq. (\ref{eq:normalization}).

\begin{equation}
    S + I + R = 1 \label{eq:normalization}
\end{equation}

The model dynamics consist of a system of 3 ordinary differential equations, presented in equations (\ref{dsdt_v}) to (\ref{drdt_v}), as shown by \cite{KeelingRohani2008}, where the dynamic begins with a determined number of infected called $I_0$ and the remainder of the population is considered susceptible. From then on, the susceptible pass to the infected group at a rate $\beta$. Infected individuals recover from the disease at a rate $\gamma$, which represents the inverse of the mean infectious period. Subsequently, the removed lose immunity at rate $\delta$. The immunity loss rate reflects both pathogen mutations that render acquired immunity ineffective \cite{SchmolkeEvasion2010, markov2023evolution} and the natural decay of antibody concentrations in the blood \cite{gaebler2021evolution}. During the vaccination campaign, the susceptible turn into removed at a constant rate $\alpha$, acquiring the same immunity as the ones who recovered from the infection. The length of the vaccination campaign is determined by the total number of vaccines and the vaccination rate. The campaign occurs until the vaccine supply is depleted. The vaccination is executed homogeneously within the population, as the SIRS model does not include an age-structured population, in contrast with other works \cite{Agur1993, wei2024studying}.

\begin{equation}
    \frac{dS}{dt} = -\beta S I + \delta R - \alpha (t) \label{dsdt_v}
\end{equation}
\begin{equation}
    \frac{dI}{dt} = \beta S I - \gamma I \label{didt_v}
\end{equation}
\begin{equation}
    \frac{dR}{dt} = \gamma I - \delta R + \alpha (t) \label{drdt_v}
\end{equation}
\begin{equation}
    \alpha(t) = \left\{ \begin{array}{ccl} 
                       \alpha & \mbox{if} & t \in [t_{0}, t_{f}] \\
                          0  &  \mbox{if} & t \notin [t_{0}, t_{f}] 
    \end{array}\right. \label{alpha}
\end{equation}

It is also essential to define the Basic Reproduction Number $(R_0)$, the average number of secondary infections produced by a primary case in a completely susceptible population, as defined in \cite{KeelingRohani2008}. Since the simulations in this study assume an initial state where the whole population is considered susceptible with the exception of one infected, $R_0$ can be taken as presented in equation (\ref{def_r0}), In this context, $R_0$ reflects the ratio between the transmission rate and the recovery rate. The basic reproduction number is also a parameter that can be estimated by the analysis of the epidemic dynamic \cite{you2020estimation}, while $\beta$ is a parameter that summarizes the pathogen infectivity and the contact rate in the mean-field approximation of the model.

\begin{equation}
    R_0 = \frac{\beta}{\gamma} \label{def_r0}
\end{equation}

The SIRS Model discussed in this paper is a robust framework for studying various infectious diseases, like Pertussis, as shown by \cite{Safan2012}, Influenza, as presented by \cite{Bucyibaruta2023} and \cite{Goncalves2011}, and, most recently, COVID-19, as exposed by \cite{ElKhalifi2024}. 

Controlling the outbreak of a disease in a major susceptible population can be approached with different objectives. Initially, it is crucial to prevent the infected peak (particularly subsequent waves) from exceeding healthcare capacity, while minimizing infections and fatalities, as done by \cite{Castioni2024rebound}. This can be achieved by reducing the contact rate between people through lockdown or other non-pharmaceutical interventions \cite{Arenas2020, Maier2020}. Alternatively, mass vaccination serves as another strategy to combat an epidemic. Again, this might be done to lower deaths and contamination, albeit could also be done to extinguish the epidemic. The SIRS model here presented, unlike \cite{Hollingsworth2011, Moore2021, Wallinga2010, Wong2020}, does not include the effects of the non-pharmaceutical interventions (NPIs), meaning that the contact rate $\beta$ is held constant.

\begin{figure}[H]
    \centering
    \includegraphics[width=1\linewidth]{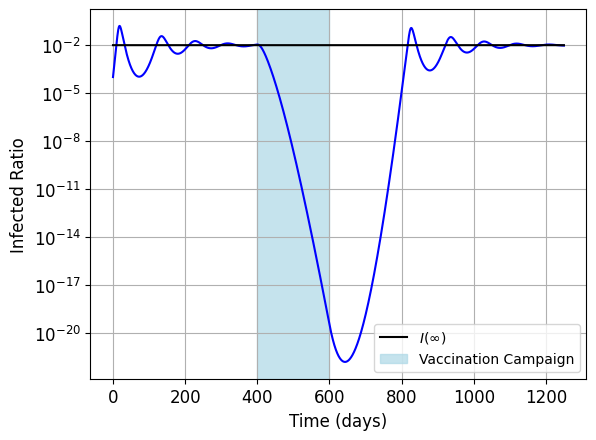}
    \caption{Time series of the SIRS Model with vaccination with $R_o = 2$, $\gamma = 0.5$, $\delta = 0.01$, $\alpha = 0.008$, $N = 10^4$ and $V = 1.6$.}
    \label{fig:deep_valley}
\end{figure}

The study of epidemic extinction within the aforementioned model is not possible because the fraction of infected can be orders of magnitude smaller than a single person, represented by the fraction $\frac{1}{N}$ in the model. This type of model can produce dynamics that diverge from real-world outcomes. For example, a scenario that would lead to an epidemic's extinction in reality might, in this model, show the epidemic persisting and even exhibiting a significant second peak as illustrated in the figure (\ref{fig:deep_valley}). 

This limitation of the continuous model is absent in agent-based models such as the one presented by \cite{Pablo2025}, where the epidemic dynamics terminate when $ I(t) = 0$. Consequently, a primary objective of this paper is to define an extinction probability and investigate its consequences for the model and the resulting epidemic dynamics. Furthermore, we examine the relationship between various parameters and the extinction probability of the epidemic, primarily the basic reproduction number $(R_0)$ and the beginning time of the vaccination campaign, to determine the optimal timing for initiating the vaccination campaign \cite{DiLauro, Costantino2019}. The possibility of delaying the vaccination campaign \cite{Maier2021_delay} for different objectives is going to be discussed along this paper. To better understand the results and to characterize epidemic extinction, the well-known analytical stationary states of the SIRS model with continuous vaccination will be used.

\section{Discussion}

\subsection{Analytic Discussion of the Model - Stationary States} \label{Analytic Discussion}

The stationary state of a system is characterized by the point where the derivatives all vanish, as shown in equation (\ref{eq:est_a_1}).

\begin{equation}
    \frac{dS}{dt}|_{S=S^{*}} = \frac{dI}{dt}|_{I=I^{*}} = \frac{dR}{dt}|_{R=R^{*}} = 0
    \label{eq:est_a_1}
\end{equation}

This section considers a vaccination campaign that differs slightly from the one presented in Equations (\ref{didt_v} - \ref{alpha}). However, the results demonstrated here will be useful throughout the remainder of this paper when analyzing the computational simulation results. In this subsection, the vaccination campaign is assumed to be continuous, beginning at a time $t=t_0$ and lasting indefinitely until the system reaches its stationary state. As demonstrated in Appendix A, the stationary state values for $S$, $I$ and $R$ are presented in Table (\ref{tab:sirs_steady_state}).

\begin{table}[h!]
\centering
\small 
\setlength{\tabcolsep}{3pt} 
\begin{tabular}{|l|c|c|c|}
\hline
\textbf{\(\alpha\)} & \textbf{\(S^{*}\)} & \textbf{\(I^{*}\)} & \textbf{\(R^{*}\)} \\ \hline \hline

\(\alpha = 0\) &
\( \frac{1}{R_0} \) &
\( \frac{\delta}{\delta+\gamma}\left(1-\frac{1}{R_0}\right) \) &
\( \frac{\gamma}{\delta+\gamma}\left(1-\frac{1}{R_0}\right) \) \\ \hline

\(0 < \alpha < \alpha_{er}\)\footnotemark &
\( \frac{1}{R_0} \) &
\( \frac{\delta}{\gamma+\delta}\left(1-\frac{1}{R_0}\right) - \frac{\alpha}{\gamma+\delta} \) &
\( \frac{\gamma}{\gamma+\delta}\left(1-\frac{1}{R_0}\right) + \frac{\alpha}{\gamma+\delta} \) \\ \hline

\(\alpha \geq \alpha_{er}\) &
\( 1 - \frac{\alpha}{\delta} \) &
0 &
\( \frac{\alpha}{\delta} \) \\ \hline

\(\alpha = \delta\) &
0 &
0 &
1 \\ \hline
\end{tabular}
\caption{Stationary state values of $S$, $I$ and $R$ in the SIRS Model with continuous vaccination campaign.}
\label{tab:sirs_steady_state} 
\end{table}

\footnotetext{The eradication vaccination rate is defined by \(\alpha_{er} = \delta\left(1-\frac{1}{R_0}\right)\).}

Vaccination rate ($\alpha$) values greater than $\delta$ also lead to total immunity within the population, although in a continuous vaccination campaign this is not physically meaningful. 

\subsection{Extinction Probability}
\label{Extinction Probability}

Based on the incongruence between the second peak following a deep trough, such as the one illustrated in Fig. (\ref{fig:deep_valley}) and the extinction of the epidemic in a discrete model or real case, the need to define an extinction probability arises. In this section, this probability is defined based on how deep the lowest valley of infected is. This definition is presented in Eq. (\ref{eq:prob_ext})

\begin{equation}
    p(m) = 
    \begin{cases}
        1 - m \cdot N, & \text{if } m < \frac{1}{N}, \\
        0, & \text{if } m \geq \frac{1}{N}.
    \end{cases}
    \label{eq:prob_ext}
\end{equation}

where p(m) is the extinction probability, m is the minimum infected fraction in a simulation and N is the community size. The idea behind this definition is to interpret an infected count of less than one individual as the probability of non-extinction. Thus, the complement of this probability is the extinction probability. When the minimum number of infected individuals exceeds one, the extinction probability is considered to be 0.

The fact that extinction probability is a function of $N$ adds a new layer of complexity to the model: the epidemic outbreak now depends on the population size. Conversely, the model presented in Eqs. (\ref{dsdt_v}) to (\ref{drdt_v}) is independent of the community size due to the normalization presented in Eq. (\ref{eq:normalization}). In Section (\ref{R0 x Extinction Probability}), will be discussed the consequences of the $N$-dependence on the epidemic extinction.

\subsection{Ensuring Model Stability and Constraints}
\label{Model corrections}

\subsubsection{Handling Susceptible Depletion}
By examining the model described in Equations (\ref{dsdt_v}) - (\ref{alpha}), one can observe that the non-negativity of the susceptible population ($S$) is not explicitly enforced. One possible solution is to impose the constraint $S > 0$ into the code, although this leads to pathological behaviors which no numerical method can solve. Therefore, inspired by \cite{Castioni2024rebound}, who terminated the simulations when S dropped below 0.005, in this paper the simulations are terminated when $S < \frac{1}{N}$. Based on the study of the stationary states for an uninterrupted vaccination campaign in Subsection \ref{Analytic Discussion}, which demonstrated that S decreasing continuously toward zero corresponds to the total immunity case, where the epidemic vanishes.  Consequently, the extinction probability is considered equal to 1.

\subsubsection{Determining the Stationary States}
Different parameter combinations lead to different epidemic dynamics, as well as varying the timescales for the system to reach the stationary state. Thus, defining a fixed time for all simulations leads to two distinct problems: either prematurely terminating simulations before the stationary state is reached, or unnecessarily extending simulations of stabilized systems for a long time,increasing the computational cost.

To solve this problem, the definition of the stationary state was employed, namely, the moment where all derivatives vanish, to find the simulation's termination point. For numerical stability, the method consists of stopping the epidemic dynamic when the euclidean norm of the derivative vector falls below a threshold value, set at $10^{-7}$. Before presenting the derivative vector, it is important to mention that the equilibrium is only achieved after the vaccination campaign has ended. Therefore, it is important to define another state variable, $V(t)$, which represents the normalized fraction of vaccinated individuals. This new state function is shown in Equation (\ref{dVdt}). Evidently, the derivative of this function is $\alpha(t)$, thus, this derivative only contributes to the euclidean norm during the campaign, as desired.

\begin{equation}
    \frac{dV}{dt} = \alpha(t) \label{dVdt}
\end{equation}

Defining $x(t) = \big(S(t), I(t), R(t), V(t)\big)$ as the derivatives vector, the euclidean norm is given by equation (\ref{euclidean norm}):

\begin{equation}
    \left\| \frac{\mathrm{d}\mathbf{x}}{\mathrm{d}t} \right\| = \sqrt{\left(\frac{\mathrm{d}S}{\mathrm{d}t}\right)^2 + \left(\frac{\mathrm{d}I}{\mathrm{d}t}\right)^2 + \left(\frac{\mathrm{d}R}{\mathrm{d}t}\right)^2 + \left(\frac{\mathrm{d}V}{\mathrm{d}t}\right)^2}
\end{equation} \label{euclidean norm}

\subsection{Comparison with the Agent-Based Model}
\label{MBA comparison}

For a new model to be valid, it must be able to explain behaviors that the previous model could not account for. Therefore, to test the validity of the proposed model and the extinction probability defined in the previous section, we present a comparison with the agent-based model proposed by \cite{Pablo2025}. 

The agent-based model proposed by \cite{Pablo2025} converges to the continuous SIRS model with vaccination as N tends to infinity; however, it exhibits extinctions due to its discrete nature. Thus, in the following section, both models are compared to evaluate the validity of the proposed model.

In the simulations of the ODE, the dynamics begin with only one infected individual and the remainder of the population is considered susceptible. In the agent-based model, such an initial state would lead to numerous stochastic extinctions that would bias the comparison of the models. For this reason, the agent dynamics are simulated starting with 10 infected individuals. This change is equivalent to shifting the infected time series forward in time and should not affect extinctions occurring later in the simulation. This ensures that the focus remains on late-stage extinctions rather than early stochastic fade-outs.

\subsection{Anomalous Epidemic Persistence at Low $R_0$ Values}
\label{R0 x Extinction Probability}

\begin{figure} [H]
    \centering
    \includegraphics[width=1.0\linewidth]{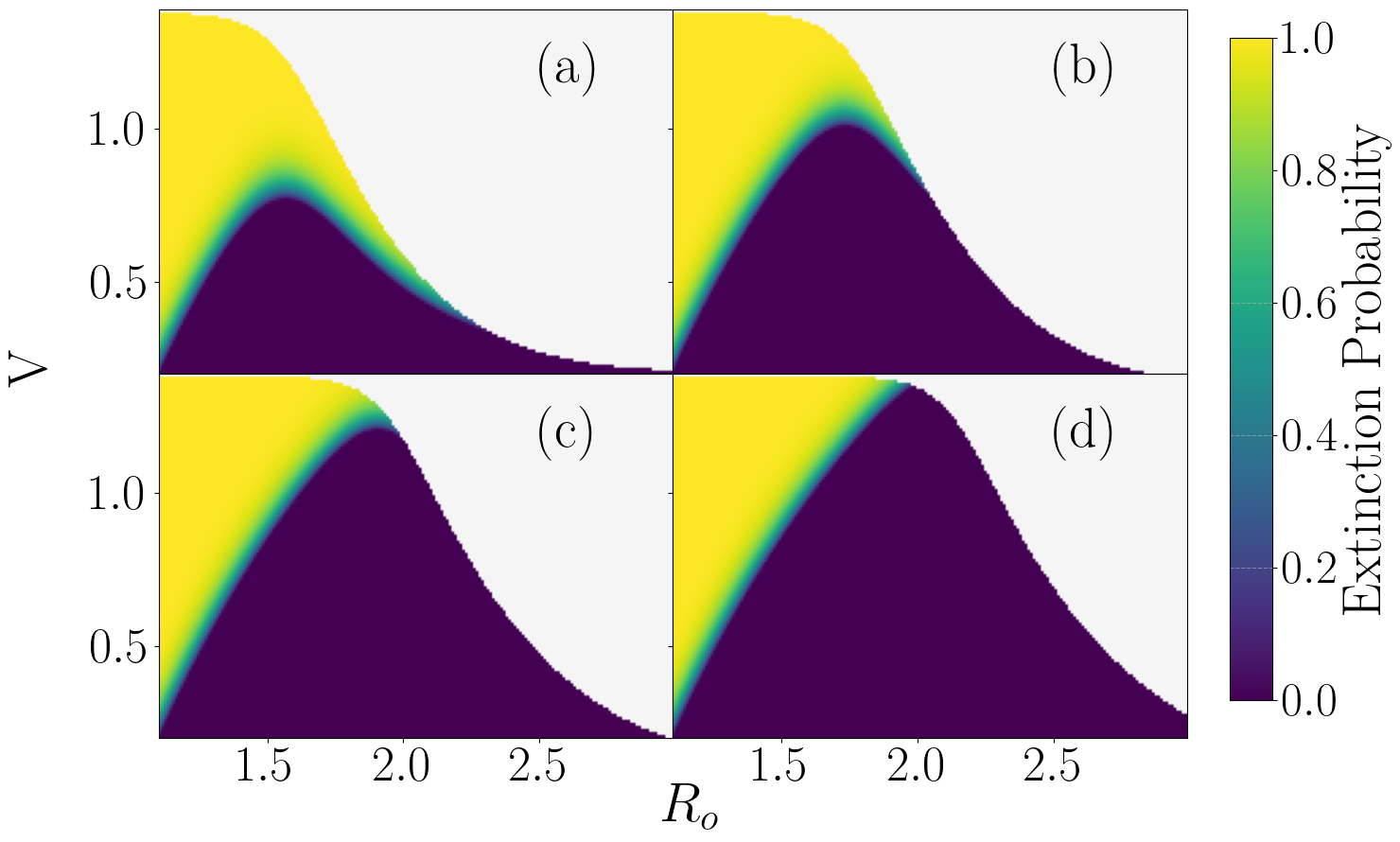}
    \caption{Extinction probability heatmap for different values of $R_0$ and vaccine counts. The gray region delimits the $R_0$ and vaccine values where the extinction probability is defined as 1 due to susceptible depletion. The population in each subplot is (a) $10^{4}$, (b) $10^{5}$, (c) $10^{6}$, (d) $10^{7}$. Fixed parameter values: $\gamma = 0.1$, $\alpha = 0.01$, $t_0 = 80$, $\delta = 0.005$.}
    \label{fig:heatmap}
\end{figure}

After preliminary analyses, an unexpected behavior of the extinction probability relative to $R_0$ was observed. Since the basic reproduction number ($R_0$) indicates the mean number of secondary infections caused by each infected individual, it is expected that a higher $R_0$ would make it more difficult to extinguish the epidemic, i.e., more vaccines would be necessary to guarantee extinction. However, this is not what is observed in Fig. (\ref{fig:heatmap}). 

As observed in Fig.~(\ref{fig:heatmap}) there is a specific value where the number of vaccines necessary to extinguish the epidemic reaches its maximum. Furthermore, the number of vaccines needed to terminate the epidemic for any given $R_0$ increases with population size, showing a rightward shift of the peak. In the context of the ordinary SIRS Model without extinction this result may be unexpected, but the epidemic persistence in bigger cities is a well known phenomenon \cite{Bartlett1957measles}, due to the lower fade-out probability.

\begin{figure}[H]
    \centering
    \includegraphics[width=\linewidth]{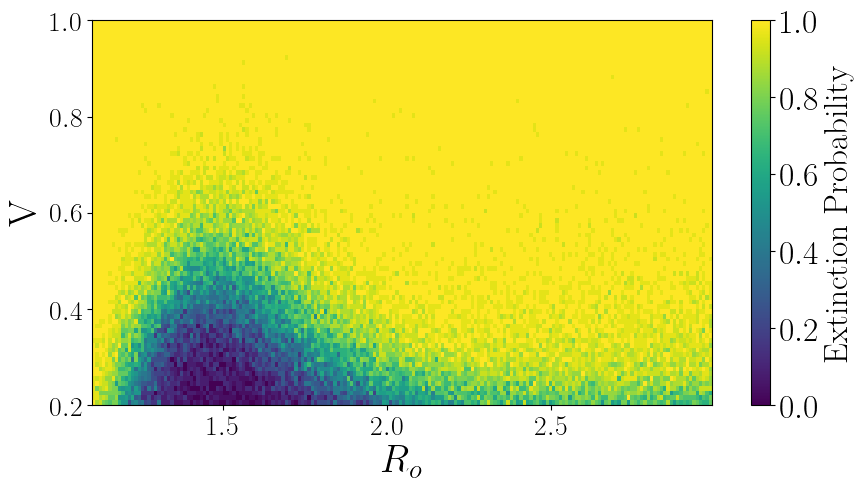}
    \caption{Extinction probability heatmap for different values of $R_0$ and number of vaccines in the agent-based Model. Fixed parameters values: $\gamma = 0.1$, $\alpha = 0.01$, $t_0 = 80$, $\delta = 0.005$ and a population of $N = 10^{4}$.}
    \label{fig:MBA VxR0}
\end{figure}

Due to this unexpected behavior, the same simulation was performed using the agent-based model presented in the previous section, maintaining the same parameters used in Fig. (\ref{fig:heatmap}). Figure (\ref{fig:MBA VxR0}) presents the result of these simulations for a population of $10^4$ agents, where the extinction probability was calculated as the ratio of the extinction cases in 25 replicates.

As shown in Fig. (\ref{fig:MBA VxR0}), the same behavior observed in the continuous model regarding a region of epidemic persistence at low values of $R_0$ is also exhibited in the agent-based model. This suggests that such behavior may be intrinsic to the SIRS model with vaccination. However, the proposed extinction probability model seems to slightly underestimate the actual probability of extinction. This is evidenced by the purple region in Fig. \ref{fig:MBA VxR0}, which is smaller than its counterpart in Fig. \ref{fig:heatmap}(a). Furthermore, the transition zone exhibits a noticeable broadening, a behavior consistent with the stochastic nature of the agent-based model. To investigate this region in greater detail, contour lines corresponding to specific extinction probabilities were added to the heatmap. Moreover, a Gaussian filter \cite[Chap. 3]{gonzalez2018digital} was applied to smooth the data and mitigate noise, facilitating visualization and enabling the generation of contour lines on the heatmap at the required values.

\begin{figure}[h]
    \centering
    \includegraphics[width=1\linewidth]{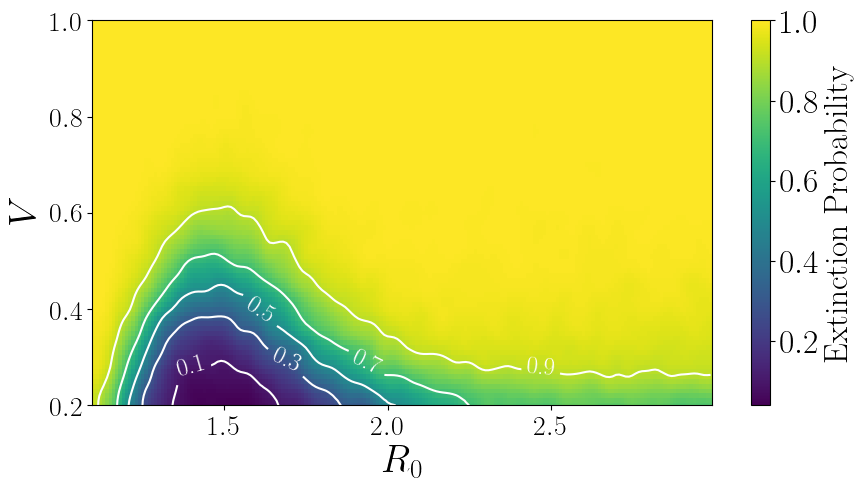}
    \caption{Extinction probability heatmap with a Gaussian filter for different values of $R_0$ and number of vaccines in the agent-based Model. Fixed parameters values: $\gamma = 0.1$, $\alpha = 0.01$, $t_0 = 80$, $\delta = 0.005$ and a population of $N = 10^{4}$.}
    \label{fig:MBA gaussian filter}
\end{figure}

The transition zone between the extinction and no-extinction areas is wider in the Agent-Based Model than in the deterministic model, which is due to its stochastic behavior. In the extinction probability definition presented in Eq.(\ref{eq:prob_ext}), the probability is non-zero only in cases where $I(t)$ becomes smaller than a single individual, while in the Agent-Based Model, stochastic extinctions may occur even in cases where the lowest infection value is greater than one individual in the deterministic model. This happens because the stochastic model presented by \cite{Pablo2025} only converges to the continuous model as $N$ goes to infinity. This is also the main reason why the proposed model underestimates the extinction probability in comparison with the  discrete model.

\begin{figure} [H]
    \centering
    \includegraphics[width=1.0\linewidth]{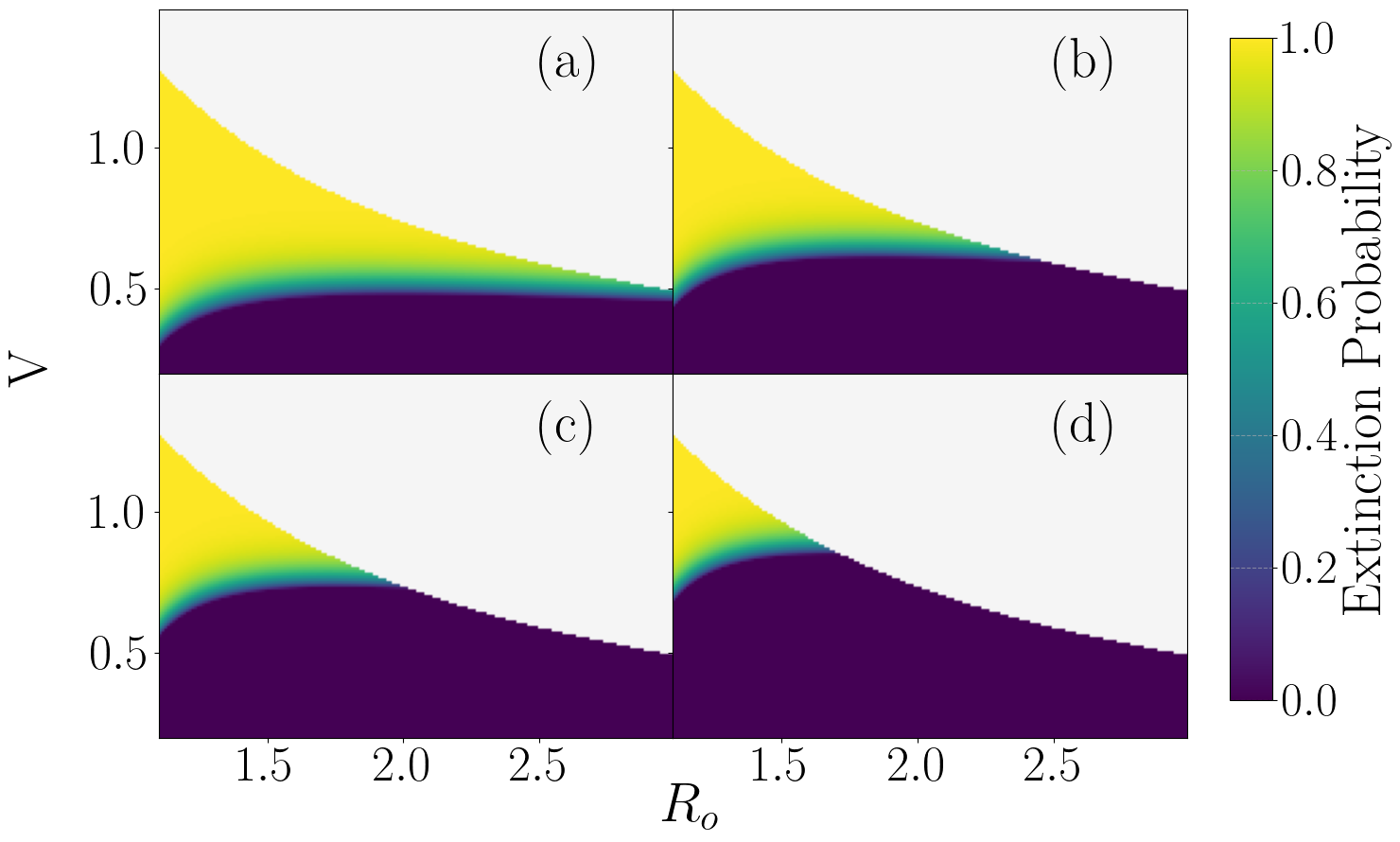}
    \caption{Extinction probability heatmap for different values of $R_0$ and number of vaccines for a vaccination campaign beginning after the stationary state. The gray region delimitates the $R_0$ and vaccine number values where the extinction probability is defined as 1 because of the lack of susceptible. The population in each subplot is (a) $10^{4}$, (b) $10^{5}$, (c) $10^{6}$, (d) $10^{7}$. Fixed parameters values: $\gamma = 0.1$, $\alpha = 0.01$, $\delta = 0.005$.}
    \label{fig:Heatmap - Stationary State - VxR0}
\end{figure}

To understand why this unexpected behavior occurs, the same simulations detailed in Fig. (\ref{fig:heatmap}) were performed, but with a key modification: vaccination was initiated only after the system reached its stationary state (without vaccination), as characterized by the $S, I,$ and $R$ values presented in Table (\ref{tab:sirs_steady_state}). The primary objective of these new simulations is to determine whether the observed behavior is dependent on the vaccination initiation time. The results of this analysis are presented in Fig. (\ref{fig:Heatmap - Stationary State - VxR0}).

By examining Fig. (\ref{fig:Heatmap - Stationary State - VxR0}), one can observe, especially in cases with smaller population sizes, a rapid expansion of the purple region at low $R_0$ values, followed by the stabilization of the transition curve. Therefore, it is possible to hypothesize that the unexpected behavior of the extinction probability is due to the timing of the vaccination campaign, as this phenomenon does not appear when vaccination is initiated at the stationary state. In the next subsection, this hypothesis will be tested and discussed in greater detail.

\subsection{Analysis of Vaccination Timing and Extinction Probability}
\label{Extinction probability maximization}

\begin{figure}[H]
    \centering

    \begin{subfigure}{\linewidth}
        \centering
        \includegraphics[width=\linewidth]{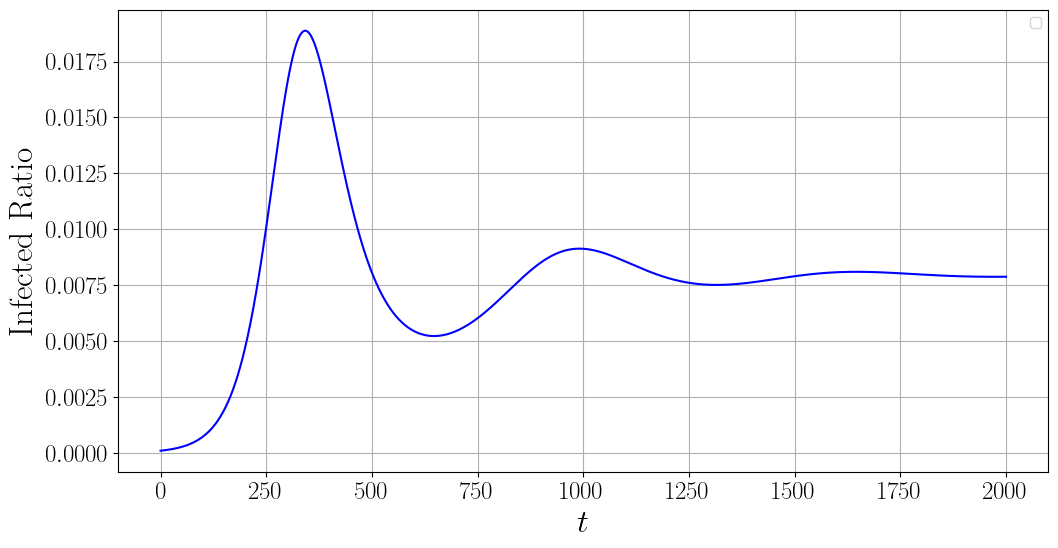} 
        \label{fig:img-a}
    \end{subfigure}%
    \hfill 
    \begin{subfigure}{\linewidth}
        \centering        \includegraphics[width=\linewidth]{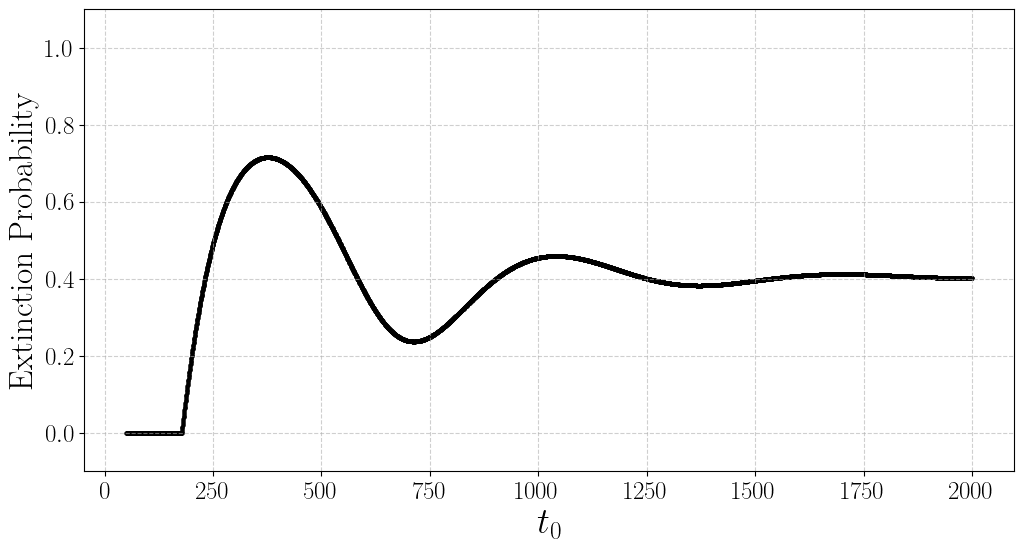} 
        \label{fig:img-b}
    \end{subfigure}%

    \caption{The top panel of this figure shows the infected time series, while the second image presents the extinction probability for each value of $t_0$. Simulation parameters: $R_0 = 1.2$, $\gamma = 0.1$, $\alpha = 0.001$, N = 10000, $\delta = 0.005$ and number of vaccines = 0.7.}
    \label{fig:Ixt and Extinction probability}
\end{figure}

As discussed in the previous section, the behavior shown in Fig. (\ref{fig:heatmap}) is likely due to the vaccination timing, in which the onset of the campaign is not optimal for specific $R_0$ values. To test this hypothesis, we provide two plots to help find the optimal time to begin the campaign for a given $R_0$, by maximizing the extinction probability. Therefore, all parameters were held fixed while the vaccination onset time is varied. Figure (\ref{fig:Ixt and Extinction probability}) presents the time series of infected in the SIRS model without vaccination, alongside the extinction probability as a function of $t_o$. 

By analyzing both plots presented in Fig.(\ref{fig:Ixt and Extinction probability}) , one can observe an analogous oscillatory behavior with a phase shift. Therefore, the need arose to estimate this phase offset between both curves. Initially, the two curves were normalized and superimposed, in order to ease the correlation analysis. Subsequently, the extinction probability curve was interpolated using the time values of the infected population curve to enable a point by point comparison. The next step involved shifting the extinction probability curve by different time intervals and computing the Pearson and Spearman correlation coefficients \cite{Schober2018}. Nevertheless, none of them yielded satisfactory results, therefore, a new method of determining the phase shift was required.

Both curves exhibit clear peaks and troughs, thus the solution found was to minimize the distance squared between corresponding extremes along the x-axis by shifting the second curve.
Namely, this involved minimizing the function f(d) defined in Eq. (\ref{d^2}), where the indices 1 and 2 refer to the infected time series and the extinction probability curves, respectively. 

\begin{equation}
    f(d) = \sum_{k} {(t_{1k} - (t_{2k} - d))^2} \label{d^2}
\end{equation}

\begin{figure} [H]
    \centering
    \includegraphics[width=\linewidth]{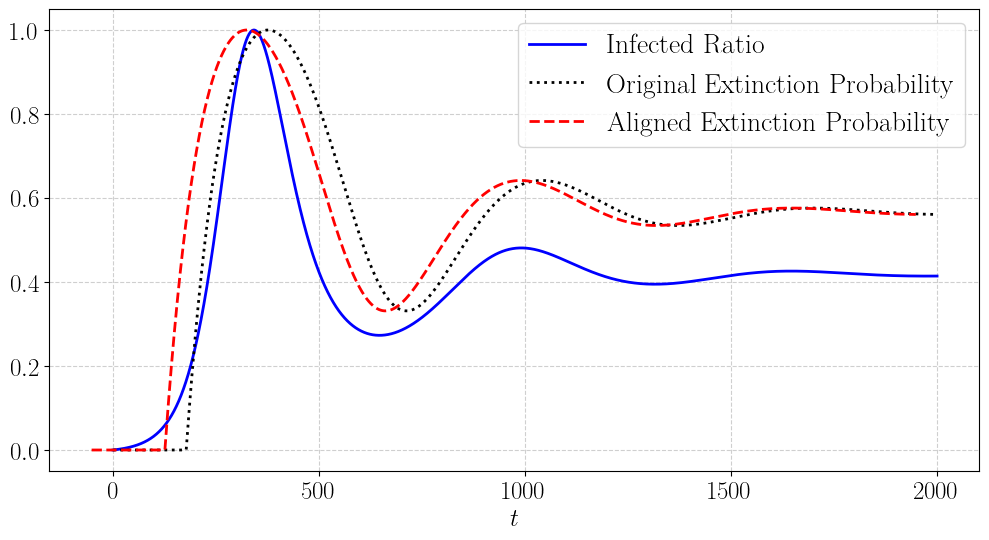}
    \caption{Comparison of the oscillatory behavior of the normalized infected series and the normalized extinction probability shifted by a distance of $d=-51.7\text{days}$.}
    \label{fig:aligned graphics}
\end{figure}

The value of $d$ that minimizes $f(d)$ corresponds to the arithmetic mean of the temporal distances between corresponding extrema, as presented in Eq.(\ref{d_minimizacao}). The result of this optimization is illustrated in Fig. (\ref{fig:aligned graphics}).

\begin{equation}
    d = \frac{\sum_{k} {(t_{2k} - t_{1k})}}{n} \label{d_minimizacao}
\end{equation}

From the results presented in Fig. (\ref{fig:aligned graphics}) it can be observed that for this set of parameters, the optimal begin time for the vaccination campaign is approximately 52 days after the first infection peak. This is consistent with the hypothesis proposed in the previous subsection. An examination of Fig. (\ref{fig:aligned graphics}) suggests that the best moment to begin the vaccination to extinguish the epidemic is after the infection peaks and close to infection troughs, thereby driving the system into deeper troughs and increasing the extinction probability. 

\begin{figure}[H]
    \centering
    \includegraphics[width=1\linewidth]{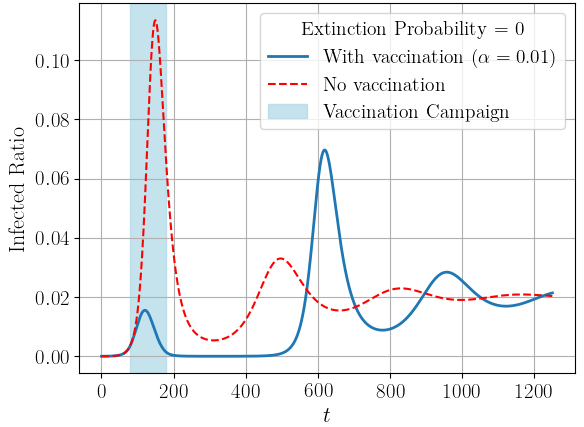}
    \caption{Comparison between the SIRS Model with and without vaccination for the parameters combination that lead to a null extinction probability. The parameters used in this simulation were $\gamma = 0.1$, $\delta = 0.005$, $R_0 = 1.74$, $\alpha = 0.01$ and number of vaccines = 1.}
    \label{fig:SIRS with and without vaccination}
\end{figure}

This does not mean that vaccination campaigns beginning before the infection peaks are necessarily a bad idea, as vaccinating the population before a peak can significantly decrease its intensity and prevent a portion of the population from becoming infected. However, vaccination during a trough apparently leads to a higher extinction probability and, therefore, is more efficient from this point of view. Consequently, campaigns beginning at a non optimal moment to extinguish the epidemic can still be useful in contexts of healthcare system overcrowding by lowering the peak of infections. Figure (\ref{fig:SIRS with and without vaccination}) illustrates a comparison between the dynamics with and without vaccination using the parameters of the highest purple point in heatmap (b) in Fig. (\ref{fig:heatmap}).

\vspace{-0.25cm}

By analyzing Fig.(\ref{fig:SIRS with and without vaccination}) one might conclude that the vaccination campaign was counterproductive because of the intense second peak emerging after the campaign, which is far greater than the second peak without vaccination. 
To address this issue, we introduce the method of the normalized integral difference between both curves, with and without vaccination, as presented by \cite{Pablo2025}. This method consists of calculating the difference between the integrals of the two curves divided by the integral of the scenario without vaccination, as shown in Eq. (\ref{eq:Integral_difference}).

\begin{equation}
    \Delta I_{total} = \frac{\int [I(t) - I_{vac}(t)] dt}{\int I(t) dt}
    \label{eq:Integral_difference}
\end{equation}

\begin{figure} [H]
    \centering
    \includegraphics[width=1\linewidth]{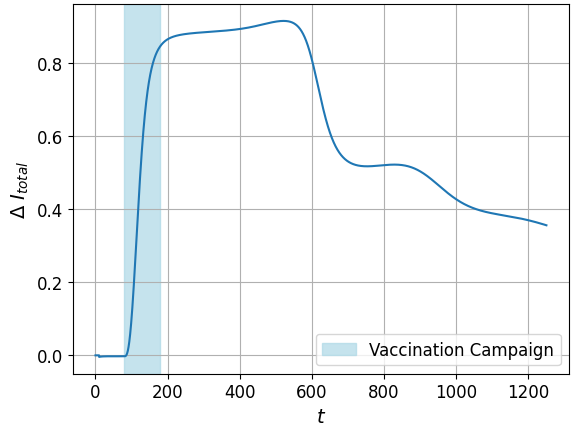}
    \caption{Graphic of $\Delta I_{total}$ between the case with and without vaccination for the following parameters: $\gamma = 0.1$, $\delta = 0.005$, $R_0 = 1.74$, $\alpha = 0.01$ and number of vaccines = 1. }
    \label{fig:Integral difference}
\end{figure}

By applying Eq. (\ref{eq:Integral_difference}) to the case shown in Fig. (\ref{fig:SIRS with and without vaccination}) and plotting the results in Fig. (\ref{fig:Integral difference}), one can observe that total number of infections throughout the entire dynamics in the vaccination scenario is lower than the SIRS model without vaccination at any time until the stationary state is reached.

\begin{figure}[h]
    \centering
    \includegraphics[width=1\linewidth]{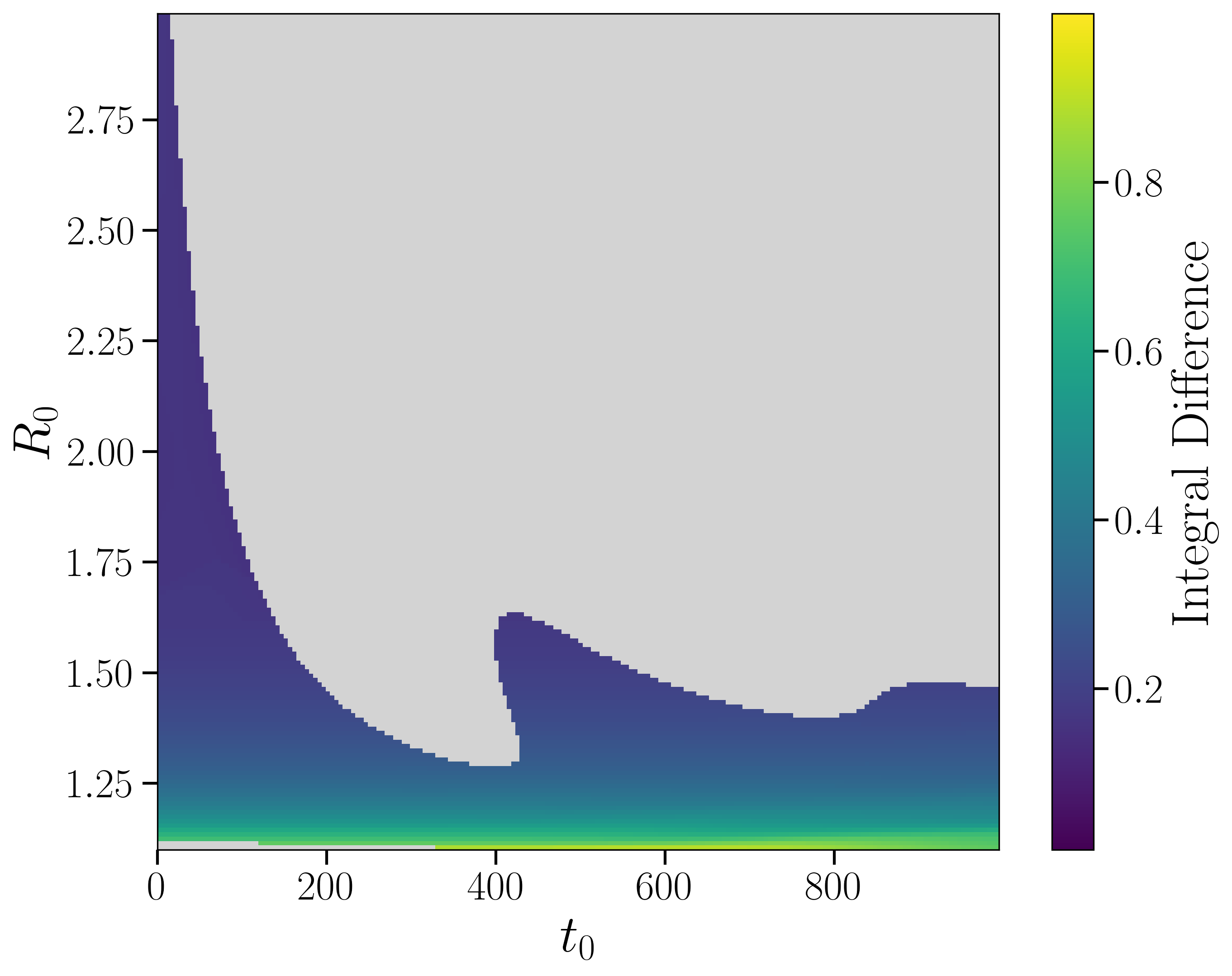}
    \caption{Graphic of $\Delta I_{total}$ between the case with and without vaccination for the following parameters: $\gamma = 0.1$, $\delta = 0.005$, $\alpha = 0.01$ and number of vaccines = 1. The light gray region corresponds to simulations where the integral difference is positive and the extinction probability is greater than 0.9.}
    \label{fig:r0 x t0}
\end{figure}

The same behavior happens for different $R_0$ values, where for all the possible values for $t_0$ the value of the integral difference defined in Eq. (\ref{eq:Integral_difference}) is positive, which means that the vaccination was beneficial for the population even though it has not begun at the optimal moment, as shown by Fig. (\ref{fig:r0 x t0}). However, as discussed before, for $R_0$ values between 1.3 and 1.8 the vaccination initiation time plays a crucial role in the extinction probability value.

\section{Summary and conclusions}

In this paper we discuss the SIRS Model with vaccination using ODEs and propose a modification to the framework to account for epidemic extinction. Since in the standard SIRS model the derivative of $I(t)$ is proportional to $I(t)$, the infection prevalence never reaches 0, but can only approach it asymptotically. Consequently, the model always predicts a secondary peak, which is inconsistent with finite populations. To address this limitation, we defined an extinction probability and analyzed the likelihood of epidemic fade-out for various parameter spaces. 

By studying the results presented in Fig. (\ref{fig:heatmap}) and Fig. (\ref{fig:Heatmap - Stationary State - VxR0}) we hypothesize that the non-monotonic behavior of the extinction probability at low $R_0$ values is driven by the vaccination initiation time. Although it can be argued that a misaligned vaccination may lead to a worse epidemic scenario, our results demonstrate that even beginning the campaign at the time where the extinction probability is minimal remains effective in terms of public health metrics. This happens due to the synchronization of the campaign with an infection peak, reducing the cumulative incidence during the epidemic, as evidenced in Fig. (\ref{fig:Integral difference}). These findings are consistent with the simulations using the agent-based model presented in \cite{Pablo2025}.

The proposal appears to transform the classic SIRS model with vaccination into a more realistic framework for scenarios in which an epidemic has a non-zero probability of extinction within a population. This is an important addition as some diseases characterized by SIRS dynamics may undergo local fade-out due to vaccination strategies. Furthermore, the modification to this framework elucidates how the system responds to immunization campaigns and why certain pathogens fail to be eradicated, remaining endemic instead.

Analysis of Fig.(\ref{fig:aligned graphics}) reveals that the optimal vaccination initiation time occurs after the infection peak. For the data presented in Fig.(\ref{fig:aligned graphics}) the ideal moment for the campaign is 52 days after the peak, however, this value is contingent upon the epidemic parameters. As shown in Fig.(\ref{fig:r0 x t0}), for some $R
_o$ values and other parameter combinations the vaccination campaign initiation time has a negligible impact on the final outcome; nevertheless, for others it may represent the difference between a successful campaign and one that fails to eradicate the epidemic.

Even though the results presented in this article are consistent with the outcomes predicted by an agent-based model, although the proposed model shares certain limitations inherent to the standard SIRS approach. Finally, we hope our work contributes to a better understanding of the impact of vaccination timing on epidemic dynamics and the conditions for pathogen extinction.

\section*{Acknowledgements}
This work was supported by Brazilian agency Conselho Nacional de Desenvolvimento Científico e Tecnológico, grants \# 309560/2025-0 and 406820/2025-2.
GH Brill and PEJ Silvestrin acknowledge CNPq for scholarship.

\appendix

\section{Demonstration of the SIRS Model Stationary States}
\label{Stationary States Demonstration}

\subsection{SIRS Model without vaccination}

The SIRS model without vaccination is the same as presented in the equations (\ref{dsdt_v}) - (\ref{drdt_v}), but considering $\alpha = 0$. 
So the following model, still normalized, is obtained:

\begin{align}
    &\frac{dS}{dt} = -\beta S I + \delta R\label{dsdt_v_A}\\
    &\frac{dI}{dt} = \beta S I - \gamma I \label{didt_v_A} \\
    &\frac{dR}{dt} = \gamma I - \delta R \label{drdt_v_A} \\
\end{align}

In the following demonstrations will be considered only the cases where $R_0$ is greater than 1, since if this condition is not fulfilled, no epidemic arises.

Based on this considerations there will be determined the values of $S^{*}$, $I^{*}$ and $R^{*}$, that correspond to the values of S, I e R in the stationary state, respectively. Using the definition of stationary state, we obtain Equation (\ref{stationary_a_1})
\begin{equation}
    \frac{dS}{dt}|_{s=s^{*}} = \frac{dI}{dt}|_{I=I^{*}} = \frac{dR}{dt}|_{R=R^{*}} = 0
    \label{stationary_a_1}
\end{equation}

Eliminating the linearly dependent equations that arise from applying (\ref{stationary_a_1}) to (\ref{dsdt_v_A}) - (\ref{drdt_v_A}) and using the normalization relation, the following equations are obtained.

\begin{align}
    &\beta S^{*}I^{*} - \gamma I^{*} = 0 \label{est_a_2}\\
    &\gamma I^{*} - \delta R^{*} = 0 \label{est_a_3}\\
    &R^{*} = 1 - S^{*} - I^{*} \label{est_a_4}
\end{align}

There are 2 possible solutions, $I^{*} = 0$ and $I^{*} \ne0$. The first case to be studied is $I^{*} \ne 0$. 
After some manipulations in (\ref{est_a_2}) and (\ref{est_a_3}):

\begin{equation}
    S^{*} = \frac{\gamma}{\beta} = \frac{1}{R_{0}} \label{est_a_5}
\end{equation}

\begin{equation}
    R^{*} = \frac{\gamma}{\delta}I^{*} \label{est_a_6}
\end{equation}

Replacing these two results in (\ref{est_a_4}):
\begin{equation}
    \frac{\gamma}{\delta}I^{*} = (1-\frac{1}{R_{0}}) - I^{*} \label{est_a_7}
\end{equation}
\begin{equation}
    I^{*} = \frac{\delta}{\delta+\gamma}(1-\frac{1}{R_{0}}) \label{est_a_8}
\end{equation}
Therefore:
\begin{equation}
    S^{*} = \frac{1}{R_{0}} \label{est_a_9}
\end{equation}
\begin{equation}
    I^{*} = \frac{\delta}{\delta+\gamma}(1-\frac{1}{R_{0}}) \label{est_a_10}
\end{equation}
\begin{equation}
    R^{*} = \frac{\gamma}{\delta+\gamma}(1-\frac{1}{R_{0}}) \label{est_a_11}
\end{equation}

\subsection{SIRS Model with a constant vaccination rate $\alpha$}
The next step is to consider an endless vaccination campaign and a vaccination rate $\alpha$. So, the stationary state equations become:

\begin{equation}
    \alpha(t) = \alpha \label{est_b_1}
\end{equation}
\begin{equation}
    -\beta S^{*}I^{*} + \delta R^{*} - \alpha = 0 \label{est_b_2}
\end{equation}
\begin{equation}
    \gamma I^{*} - \delta R^{*} + \alpha = 0 \label{est_b_3}
\end{equation}
\begin{equation}
    R^{*} = 1 - S^{*} - I^{*} \label{est_b_4}
\end{equation}

Handling equation (\ref{est_b_4}) to eliminate $R^{*}$ from the system of equations.

\begin{equation}
    \beta S^{*}I^{*} + \delta - \delta S^{*} - \delta I^{*} - \alpha = 0 \label{est_b_5}
\end{equation}
\begin{equation}
    \gamma I^{*} - \delta + \delta S^{*} + \delta I^{*} + \alpha = 0 \label{est_b_6}
\end{equation}
Adding (\ref{est_b_5}) and (\ref{est_b_6}) and rearranging algebraically, the result obtained is:
\begin{equation}
    -\beta S^{*}I^{*} + \gamma I^{*} = 0 \label{est_b_7}
\end{equation}
\begin{equation}
    S^{*} = \frac{\gamma}{\beta} = \frac{1}{R_{0}} \label{est_b_8}
\end{equation}
Substituting (\ref{est_b_8}) in (\ref{est_b_5}), we obtain:
\begin{equation}
    -\beta \frac{\gamma}{\beta} I^{*} + \delta - \frac{\delta\gamma}{\beta} - \delta I^{*} - \alpha = 0 \label{est_b_9}
\end{equation}
\begin{equation}
    \delta(\frac{\beta-\gamma}{\beta}) - \alpha = I^{*}(\gamma+\delta) \label{est_b_10}
\end{equation}
\begin{equation}
    I^{*} = \frac{\delta}{(\gamma+\delta)}(\frac{\beta-\gamma}{\beta}) - \frac{\alpha}{\gamma+\delta} \label{est_b_11}
\end{equation}
\begin{equation}
    I^{*} = \frac{\delta}{\gamma+\delta}(1-\frac{1}{R_{0}}) - \frac{\alpha}{\gamma+\delta} \label{est_b_12}
\end{equation}
Calculating $R^{*}$ by substituting $I^{*}$ and $S^{*}$ in (\ref{est_b_4}):
\begin{equation}
    R^{*} = (1-\frac{1}{R_{0}}) - \frac{\delta}{\gamma+\delta}(1-\frac{1}{R_{0}}) + \frac{\alpha}{\gamma+\delta}
\end{equation}
\begin{equation}
    R^{*} = (1-\frac{\delta}{\gamma+\delta})(1-\frac{1}{R_{0}}) + \frac{\alpha}{\gamma+\delta}
\end{equation}
\begin{equation}
    R^{*} = \frac{\gamma}{\gamma+\delta}(1-\frac{1}{R_{0}}) + \frac{\alpha}{\gamma+\delta}
\end{equation}
For $I^ > 0$, it follows that:
\begin{equation}
    I^{*} = \frac{\delta}{\gamma+\delta}(1-\frac{1}{R_{0}}) - \frac{\alpha}{\gamma+\delta} > 0
\end{equation}
Therefore:
\begin{equation}
    \alpha_{co} < \alpha_{er} = \delta(1-\frac{1}{R_{0}})
\end{equation}
Where $\alpha_{er}$ is known as the eradication vaccination rate, since for this vaccination rate the epidemic goes extinct and $\alpha_{co}$ is the coexistence vaccination rate, where the epidemic reaches the endemic state.
The next case to be analyzed is when $\alpha \geq \alpha_{er}$, for which $I^{*}=0$. Substituting $I^{*}$ into the stationary state equations:
\begin{equation}
    \delta R^{*} - \alpha = 0 
\end{equation}
\begin{equation}
    R^{*} = 1 - S^{*} 
\end{equation}
\begin{equation}
    R^{*} = \frac{\alpha}{\delta}
\end{equation}
\begin{equation}
    S^{*} = 1 - \frac{\alpha}{\delta}
\end{equation}
In the case where $\alpha = \delta$, $S^{*}$ becomes 0 and all the population is immune to the disease. This $\alpha$ value is known as the total immunity $\alpha$.

\bibliographystyle{elsarticle-harv} 
\bibliography{example}






\end{document}